\documentclass[fleqn,usenatbib]{mnras}

\usepackage[T1]{fontenc}

\DeclareRobustCommand{\VAN}[3]{#2}
\let\VANthebibliography\thebibliography
\def\thebibliography{\DeclareRobustCommand{\VAN}[3]{##3}\VANthebibliography}

\usepackage[pdftex]{graphicx}	
\usepackage{amsmath}	
\usepackage{amssymb}	
\usepackage{epsfig,graphics,subfigure,psfrag,amsmath,amssymb,amscd}
\usepackage{float}

\title[AT2018dyk Revisited]{AT2018dyk Revisited: a Tidal Disruption Event Candidate with Prominent Infrared Echo and Delayed X-ray Emission in a LINER Galaxy}

\author[Huang et al.]{Shifeng Huang$^{1,2}$\thanks{E-mail: sfhuang999@ustc.edu.cn}, Ning Jiang$^{1,2}$\thanks{E-mail: jnac@ustc.edu.cn}, Zheyu Lin$^{1,2}$, Jiazheng Zhu$^{1,2}$, Tinggui Wang$^{1,2}$
\\
\\
$^1$CAS Key Laboratory for Research in Galaxies and Cosmology, Department of Astronomy, University of Science and Technology of China, \\
Hefei, 230026, China\\
$^2$School of Astronomy and Space Sciences, 
University of Science and Technology of China, Hefei, 230026, China}

\begin{document}

\maketitle

\begin{abstract}
The multiwavelength data of nuclear transient AT2018dyk, initially discovered as a changing-look low-ionization nuclear emission-line region (LINER) galaxy, has been revisited by us and found being in agreement with a tidal disruption event (TDE) scenario. The optical light curve of AT2018dyk declines as a power-law form approximately with index -5/3 yet its X-ray emission lags behind the optical peak by $\sim140$ days, both of which are typical characteristics for TDEs. The X-ray spectra are softer than normal active galactic nuclei (AGNs) although they show a slight trend of hardening. Interestingly, its rising time scale belongs to the longest among TDEs while it is nicely consistent with the theoretical prediction from its relatively large supermassive black hole (SMBH) mass ($\sim10^{7.38} M_{\sun}$). Moreover, a prominent infrared echo with peak luminosity $\sim7.4\times10^{42}$~\text{erg}~$\text{s}^{-1}$ has been also detected in AT2018dyk, implying an unusually dusty subparsec nuclear environment in contrast with other TDEs. In our sample, LINERs share similar covering factors with AGNs, which indicates the existence of the dusty torus in these objects. Our work suggests that the nature of nuclear transients in LINERs needs to be carefully identified and their infrared echoes offer us a unique opportunity for exploring the environment of SMBHs at low accretion rate, which has been so far poorly explored but is crucial for understanding the SMBH activity.
\end{abstract}


\section{Introduction} \label{sec:intro}
In recent years, more and more transients have been discovered in the nuclear of galaxies as various sky surveys have been conducted. Despite rapid growth in their number, the origins of these transients are still under debate. Due to the presence of supermassive black holes (SMBHs) at the centers of galaxies, some of these sources are likely to be related to the accretion process of the central black hole. 

Among the nuclear transients found so far, although supernovas dominate the population, and there are still a substantial fraction of them are ascribed to tidal disruption events (TDEs), the activity of active galactic nuclei (AGNs), and the still unidentified ambiguous nuclear transients (ANTs). TDEs are phenomena in which the tidal force of a black hole at the center of a galaxy tears apart a star \citep{hills1975,rees1988}. Most optical TDEs exhibit a blue continuum and broad emission lines (e.g., H$\alpha$ and He \textsc{ii}) in their spectra, a black body component with a slowly evolving temperature of a few$\,\times10^4$ K in the spectral energy distribution (SED); and a power-law decline in the light curves \citep{arcavi2014,velzen2020,gezari2021,zabludoff2021}. Interestingly, some ANTs show the above characteristics and also exhibit some features in AGNs \citep{hinkle2022}. Most of the current TDEs are found in non-active galaxies, and it is a difficult task to identify the cause of the transients occurring in AGNs. Nevertheless, TDEs have been found in some AGNs, such as PS16dtm \citep{blanchard2017,jiang2017}, CSS100217:102913 + 404220 \citep{cannizzaro2022}, CSS J102913+404220 \citep{zhangw2022} and even in the blazar OJ 287, the possible TDE-induced outbursts are found \citep{huang2021,huang2022}.

AGNs are the most luminous objects in the universe, and the unified model suggests that they possess structures such as dusty tori and accretion disks outside the black holes \citep{urry1995}. The obscured fraction of galaxies can be described by the dust's covering factor, which decreases as the luminosity rises due to obscuring material being blown away \citep{hickox2018}. Additionally, the degree of obscuration depends on the Eddington ratio, implying that the dusty torus is related to radiation pressure \citep{ricci2017b}. Low-ionization nuclear emission-line regions (LINERs) are a subclass of AGN accounting for about 2/3 of AGNs and 1/3 of nearby galaxies \citep{ho2008}. These objects are in the low state of AGN evolution with low accretion rates and luminosities \citep{falcke2004}. Despite what is known about LINERs, there has been controversy surrounding the existence of dusty tori in such objects. \cite{martin2015} detected dusty tori in some LINERs, while \cite{balmaverde2015} found no evidence of such structures in their sample. \cite{almeida2017} suggested that this may be due to the disappearance of the torus structure in low-luminosity AGNs.

When photons of X-ray, UV, and optical wavelengths are radiated from the center of a galaxy into surrounding dust, the emissions are absorbed and converted into infrared (IR) radiation \citep{mattila2018,reynolds2022}. The detection of IR echoes from transients in the galactic nucleus can therefore be a tool to identify the presence of dust. IR echo signals have been observed after some TDEs, such as ASASSN-14li \citep{jiang2016,velzen2016}. Additionally, \cite{jiang2021b} analyzed IR echoes from 23 optical TDEs and found that the covering factor of dust in host galaxies of these TDEs is much lower than that of AGNs. Furthermore, \cite{hinkle2022} analyzed IR echo signals from some ANTs and obtained their covering factors.

AT2018dyk was discovered by the Zwicky Transient Facility (ZTF) on 31 May 2018 with a magnitude of 19.41 in the $g$ band with the host galaxy at a redshift of 0.0367 \citep{frederick2019}. This nuclear transient was classified as a TDE due to its blue continuum and broad emission lines of H$\alpha$, H$\beta$, and He \textsc{ii} \citep{arcavi2018}. However, \cite{frederick2019} found that after AT2018dyk was detected, its host galaxy transitioned from a LINER to a narrow-line Seyfert 1 galaxy, which may have been caused by activity on the accretion disk. \citet{hinkle2021} fitted a black body model to the spectral energy distribution in UV/optical bands, and the results showed a slowly evolving temperature of $\sim10^4$ K. Additionally, the evolution of the black body radius is similar to that of TDEs \citep{holoien2022}. Recently, \cite{hinkle2022} studied the IR data and derived a dust covering factor of $0.42\pm{0.15}$.

In this work, we perform a multiwavelength analysis of AT2018dyk, including the X-ray, UV, optical, and IR bands, to explore the nature of this intriguing LINER nuclear transient and its environment. In Section \ref{sec:analysis}, we describe the data analysis and present results from multiwavelength light curves and X-ray spectra. Next, in Section \ref{sec:discussion}, we discuss the physical origin of AT2018dyk. Finally, we conclude this work in Section \ref{sec:conclusion}. Throughout this work, we assume cosmological parameters of $H_0=70\text{km}\text{s}^{-1},\text{Mpc}^{-1}$, $\Omega_{\rm M}=0.3$, and $\Omega_{\Lambda}=0.7$.

\section{Data Analysis and Results}\label{sec:analysis}

\subsection{X-ray Data}
The Swift X-Ray Telescope (XRT) is equipped on the Neil Gehrels Swift Observatory \citep{burrows2005}. We download the public data of 27 observations from July 2018 to February 2021. Additionally, we requested a Target of Opportunity (ToO) observation on December 5, 2022 (Observation ID: 00010780020, PI: Huang). With \texttt{HEASoft 6.30.1}, we reduce the data using \texttt{xrtpipeline} to obtain the level 2 files. Light curves and spectra are derived by \texttt{xrtproducts} using a source region with a radius of $35^{\prime\prime}$ and a source-free background region with a radius of $100^{\prime\prime}$. 

X-ray Multi-Mirror Mission (XMM-Newton) is an X-ray observatory of the European Space Agency, and the European Photon Imaging Camera (EPIC) is one of the main instruments, including two MOS detectors and a pn camera. We obtain the data in two observations (obs IDs: 0822040701, PI: Gezari; and 0865051001, PI: Frederick) by XMM-Newton, which are reduced by Science Analysis System (SAS version 20.0). Following the recommended data analysis threads, tasks \texttt{cifbuild} and \texttt{odfingest} are executed for the preparation, and then the light curves and spectra are extracted by \texttt{xmmextractor}.  
 
The first observation by Swift/XRT was performed in MJD 58330.  Assuming the absorbed power-law spectrum with an index of $\Gamma=1.75$ \citep{ricci2017} and a Galactic hydrogen density of $1.59\times 10^{20}\,\text{cm}^{-2}$ \citep{HI4PI2016}, we derive the 3$\sigma$ upper limit of $5.07\times 10^{-13}~\text{erg}~\text{s}^{-1}~\text{cm}^{-2}$ for the flux in 0.3 -- 10.0 keV using WebPIMMS\footnote{\url{https://heasarc.gsfc.nasa.gov/cgi-bin/Tools/w3pimms/w3pimms.pl}}. The XMM-Newton observed the source 12 days later. Fixing the neutral hydrogen column density, the X-ray spectra were analyzed through \texttt{xspec 12.12.1} with the model $\texttt{tbabs*zashift*powerlaw}$. We derived a flux of $(2.87^{+0.36}_{-0.23})\times 10^{-13}~\text{erg}~\text{s}^{-1}~\text{cm}^{-2}$ in 0.3 -- 10.0 keV and $\Gamma=3.04^{+0.33}_{-0.30}$ (C/d.o.f = 14.03/15) by fitting the EPIC-pn spectrum with an absorbed power-law model. The observed peak luminosity of $(4.21^{+0.52}_{-0.44})\times~10^{42}~\text{erg}~\text{s}^{-1}$ was detected in MJD 58460. During the peak, fitting with a single power-law, we obtain the spectral index of $\Gamma=2.47^{+0.35}_{-0.34}$ (C/d.o.f = 9.72/12). The X-ray luminosity stayed at the plateau for a long time until MJD 58753. After that, the source declined and in MJD 59020, an upper limit of $3.31\times 10^{-14}~\text{erg}~\text{s}^{-1}~\text{cm}^{-2}$ was obtained by EPIC-pn. From then on, the source has been an X-ray faint source and in the last observation in MJD 59917, only an upper limit of $8.62\times 10^{-14}~\text{erg}~\text{s}^{-1}~\text{cm}^{-2}$ was obtained by Swift/XRT with the exposure time of 2688 s.

\begin{table}
\centering
\caption{For different epochs, the X-ray spectra were fitted by an absorbed power-law model. Here, we fixed the hydrogen density as $1.59\times 10^{20}\,\text{cm}^{-2}$. In the second column, $\Gamma$ denotes the photon index and the results tested by Cash statistic are shown in the third column. }
\begin{tabular}{lcc}
\hline\\
 \multicolumn{1}{c}{Spectrum} & \multicolumn{1}{c}{$\Gamma$} & \multicolumn{1}{c}{C/d.o.f} \\ \hline
 Initial detection &  $3.04^{+0.33}_{-0.30}$ & 14.03/15\\
  Rising stage stacked & $2.73^{+0.16}_{-0.16}$ & 59.92/50\\
  Peak  & $2.47^{+0.35}_{-0.34}$ & 9.72/12\\
  Plateau stage stacked & $2.20^{+0.15}_{-0.14}$ & 59.31/48\\ \hline
\end{tabular}
\label{tab:spec_fit}
\end{table}

\begin{figure*}
  \centering
  \subfigure[]{
  \includegraphics[width=0.45\textwidth]{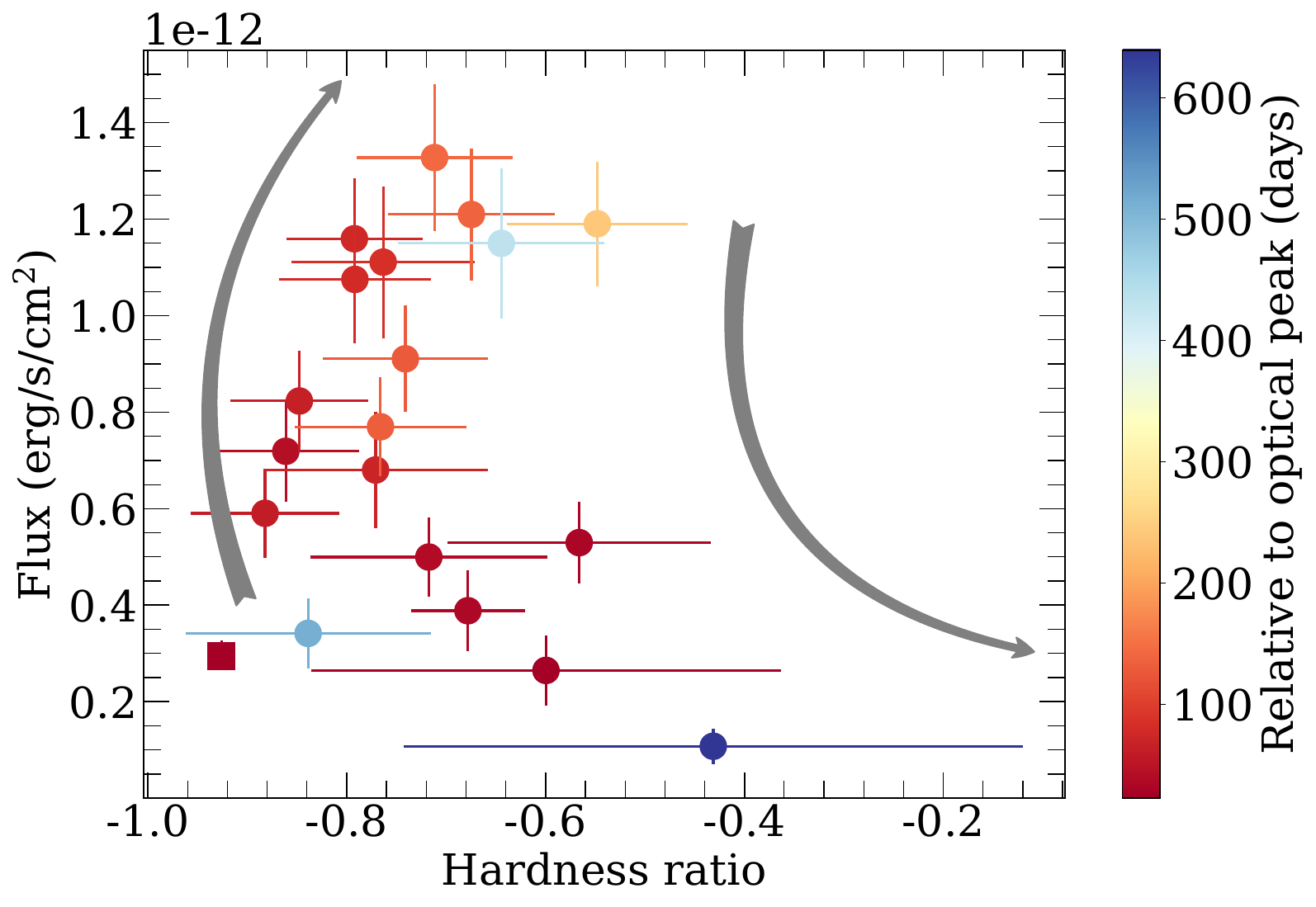}}
  \subfigure[]{
  \includegraphics[width=0.45\textwidth]{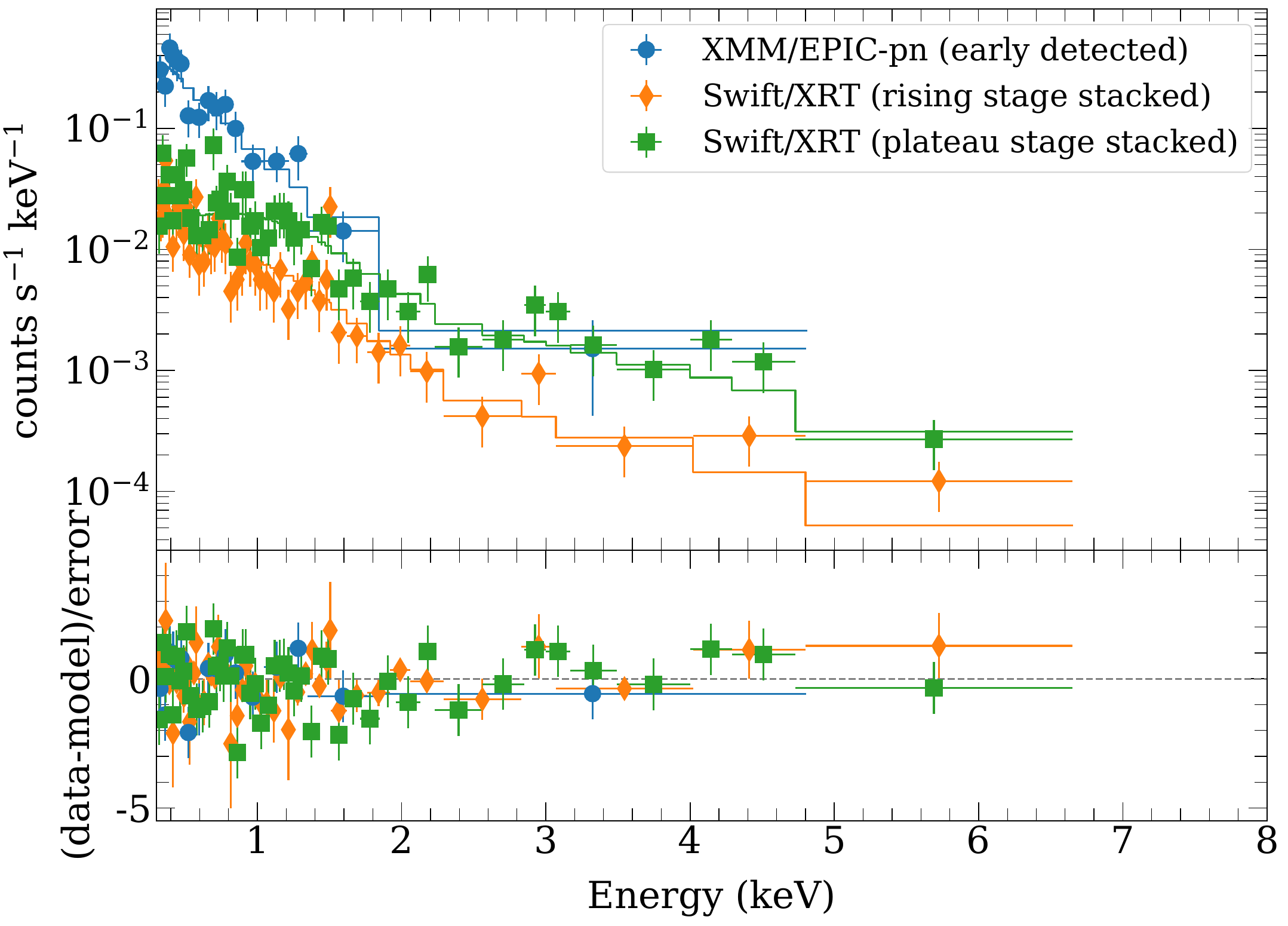}}
  \caption{The X-ray evolution. In panel (a), the circles and squares represent Swift/XRT and XMM-Newton observations, respectively. The curve arrows 
  mark the overall evolution tendency of the HR. In panel (b), the X-ray spectra from initial detection to the peak are plotted.}\label{fig:hr}
\end{figure*}

The hardness ratio is defined as HR$\,\equiv(H-S)/(H+S)$, where $H$ and $S$ represent the net count rates for hard (2.0 -- 10.0 keV) and soft (0.3 -- 2.0 keV) X-ray bands, respectively. The evolution of HR with flux is displayed in panel (a) of Figure~\ref{fig:hr}. The soft X-ray dominated the flux in the initial observations. However, as the flux rose, the HR gradually increased. After the plateau stage, the HR increased when the source got fainter. Fitted by a single absorbed power-law, the stacked spectra in the rising stage and plateau stage arrive at the spectral indexes of $2.73^{+0.16}_{-0.16}$ (C/d.o.f = 59.92/50) and $2.20^{+0.15}_{-0.14}$ (C/d.o.f = 59.31/48), respectively. In spite of the hardening trend during the X-ray brightening stage, it is still softer than the case in typical AGNs. The X-ray spectral fitted results in different epochs are listed in Table \ref{tab:spec_fit}. 

\subsection{UV/Optical Data}
The UV/Optical Telescope (UVOT) is one of the instruments on the Swift Observatory \citep{roming2005}. We first examine each image file and exclude the extensions with bad photometric flags. For image files with multiple valid extensions, we sum all extensions using the task \texttt{uvotimsum}. And then, the task \texttt{uvotsource} performs photometry on each image, with the source and source-free background region defined by a circle of the radius of $5^{\prime\prime}$ and $50^{\prime\prime}$, respectively. As the magnitudes have remained unchanged in the past two years, we take the latest measurement as the contribution of the host galaxy, and a subtraction process results in differential magnitudes. Following \cite{schlafly2011}, the Galactic extinction of $E(B-V)=0.0164$ is derived by an online tool\footnote{\url{https://irsa.ipac.caltech.edu/applications/DUST/}}, and the magnitude for each band is corrected by using the extinction law of \cite{cardelli1989}.

We obtain the point spread function light curves of AT2018dyk through the ZTF forced-photometry service \citep{masci2019}. We filter out the photometry results that are impacted by bad pixels or bad seeing, and then we perform baseline corrections for AT2018dyk based on the quiescent state before the outburst. All photometry are performed on the differential images taken in the same field, charge-coupled device (CCD), and CCD-quadrant. 

\begin{figure*}
  \centering
  \includegraphics[width=0.9\textwidth]{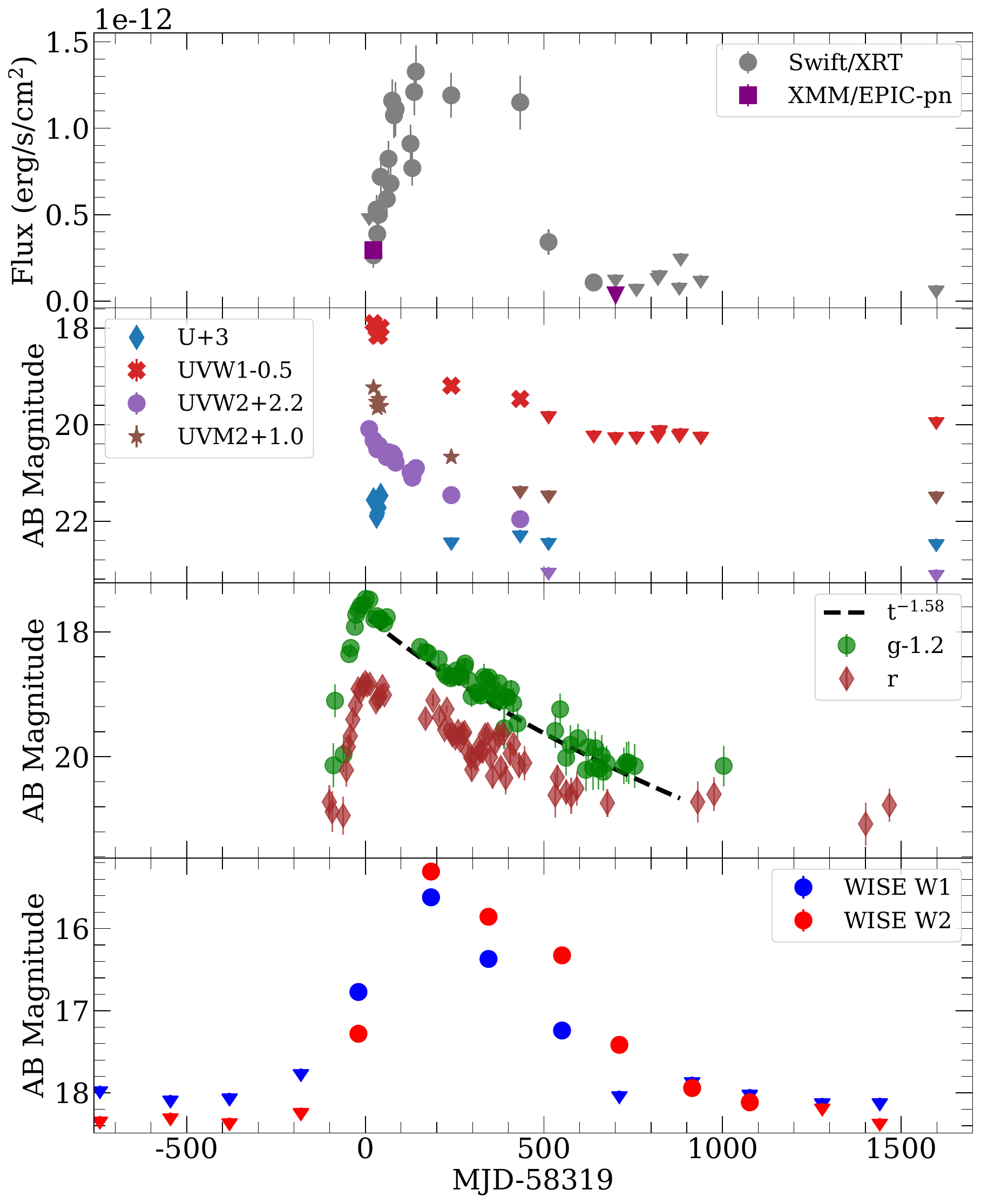}
  \caption{Multiwavelength light curves of AT2018dyk. From top to the bottom, the X-ray (0.3--10.0 keV), UV, optical, and IR light curves are plotted. The UV/optical and IR magnitudes are host-subtracted. Owing to the low signal-to-noise after host-subtraction, the data points in V and B bands are not shown here. Data points in the g and r bands are binned in 7 days in the third panel. Additionally, the dashed line in the third panel represents the power-law fitting results for the $g$ band luminosity. In each panel, the triangles represent the 3$\sigma$ upper limit. }\label{fig:mwlc}
\end{figure*}

\begin{figure}
  \centering
  \includegraphics[width=0.45\textwidth]{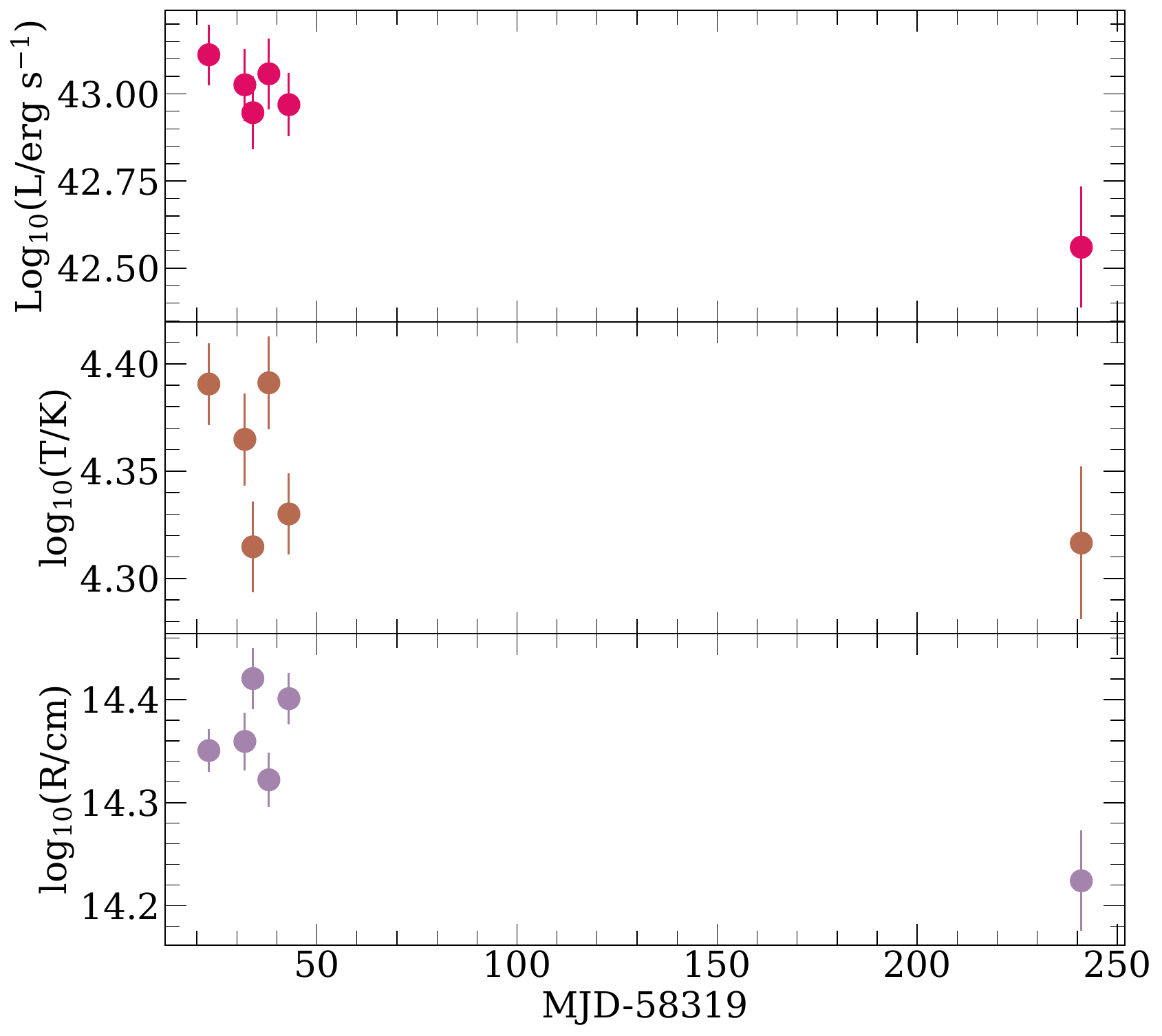}
  \caption{From the top to the bottom, is the evolution of UV/optical luminosity, black body temperature, and radius, respectively. }\label{fig:disk_temper}
\end{figure}

AT2018dyk was discovered by ZTF\footnote{\url{https://www.wis-tns.org/object/2018dyk}} on 31 May 2018, with an AB magnitude of 19.4. After that, about 2 months later, its optical brightness reached a peak value of 18.7 mag which was observed in around MJD 58319. And then, it declined slowly to 21.4 mag in MJD 59322. Unfortunately, Swift/UVOT missed the whole rising stage. The observed maximum brightness in UV bands was detected at 18.3 mag in the UVW2 band in MJD 58330. After that, the brightness in UV/optical bands continuously declined. In the latest observation by Swift/UVOT, magnitudes of the whole galaxy in V, B, U, UVW1, UVM2, and UVW2 bands are 15.53, 16.25, 17.3, 17.81, 17.85, and 17.66 mag, respectively. The multiwavelength light curves are shown in Figure~\ref{fig:mwlc}. The host-subtracted light curve of the $g$ band is fitted by $F=A~t^{-\alpha} + h$, where $F$ is the flux, $A$ and $h$ are constants, and the best fitted value of $\alpha$ is 1.58. The UV/optical SEDs are fitted by a black body model and we obtain the evolution of temperature and black body radius in Figure~\ref{fig:disk_temper}. It is worth mentioning that the blackbody temperatures of AT2018dyk are quite high during the outbursts and there is no significant tendency to decrease with decreasing luminosity. This result is consistent with \cite{hinkle2021} and \cite{holoien2022}.

\subsection{Mid-infrared Data}
AT2018dyk has been continuously observed by Near-Earth Object Wide-field Infrared Survey Explorer (NEOWISE, \citealt{Mainzer2014}), which is a successor of WISE~\citep{Wright2010}, at W1 (3.4~$\mu$m) and W2 ($4.6\mu$m) bands every six months since 2014 January. The public NEOWISE catalog has offered us
a total of 16 visits up to 2021 June with typically 12 individual exposures in each epoch. Its W1 and W2 magnitudes keep constant until a sudden rising since 2018 June (MJD=58230) and then reach to peak immediately at the next epoch, which is followed by a steady declining to the constant level. Before the outburst, the average magnitude of the galaxy in the W1 and W2 bands were both $\sim 12~\text{mag}$, and $\text{W1}-\text{W2}<0.04~\text{mag}$. The observed peak of IR bands was detected in MJD 58503, lagging the optical peak for $\sim 180$ days, at 11.6 mag and 11.26 mag for W1 and W2 bands, respectively. It should be noted that during the IR outburst, $\text{W1}-\text{W2}\gtrsim 0.1~\text{mag}$, and after the outburst, $\text{W1}-\text{W2}< 0.04~\text{mag}$ again. This indicates a heating and cooling process, which will be discussed in Section \ref{sec:dust}.

\section{Discussion}\label{sec:discussion}
AT2018dyk was reported by \cite{frederick2019} as a `changing-look' phenomenon that turned on and a transition from the LINER to narrow-line Seyfert 1 was observed. They interpreted the transition as caused by the physical activity on the accretion disk. The IR observations reveal the existence of the dusty torus in this system and through the IR data, we can obtain some information about the dust. There is still controversy about the existence of dusty torus in LINERs. Here, we analyse the covering factors of some of the galaxies where the nuclear transients are located and found that the covering factors of LINERs in the sample are not significantly different from those of AGNs. In this section, based on the analysis of the light curves and X-ray spectra, we infer that the transient nuclear outburst is probably induced by a TDE.

\subsection{The dust in the system}\label{sec:dust}
\begin{figure}
  \centering
  \includegraphics[width=0.45\textwidth]{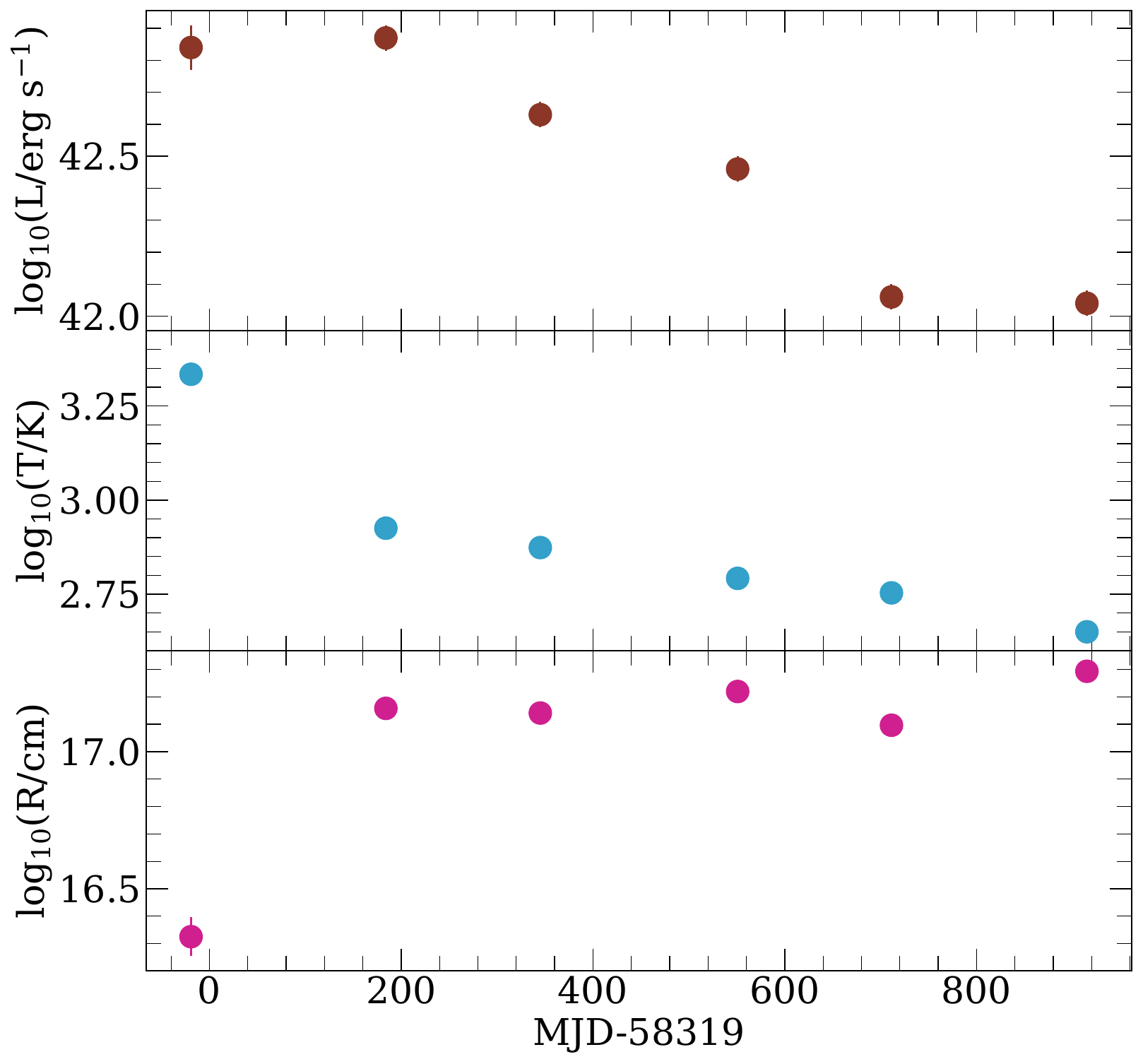}
  \caption{The evolution of dust properties. The upper panel plots the decline of the IR luminosity, while the evolution of dust temperature and radius are plotted in the middle and bottom panel.}\label{fig:dust}
\end{figure}

When the UV/optical outburst caused by the transient accretion onto SMBHs occurs, the surrounding dust will unavoidably absorbs part of the emission and transforms it into IR emission through reprocessing. Thus, the dust IR echoes provide us an efficient way to probe the dust around SMBHs, i.e. the dusty torus in AGNs. 
After subtracting the IR emission before the outburst, which is dominated by host galaxy, we obtain the properties of heated dust  in each epoch with blackbody SED fitting (see Figure~\ref{fig:dust}). At the IR peak (around MJD 58503), we derive a luminosity of $(7.41\pm{0.68})\times10^{42}~\text{erg}~\text{s}^{-1}$ and a temperature of $842.07\pm{16.81}~\text{K}$. It should be noted that before the IR peak, the dust temperature is even higher, that is $2158.39\pm{154.04}~\text{K}$ indicating a likely dust sublimation process in the early stage after the outburst~\citep{jiang2017} due to the proximity of the dust around the SMBH. As the IR luminosity declines, the dust cools down while the radius remains at a relatively flat level. We also estimate a dust covering factor of $0.40\pm{0.14}$ using the ratio of the peak luminosity of optical to IR bands \citep{jiang2021}, and the result is consistent with \cite{hinkle2022}.

\begin{figure}
  \centering
  \includegraphics[width=0.45\textwidth]{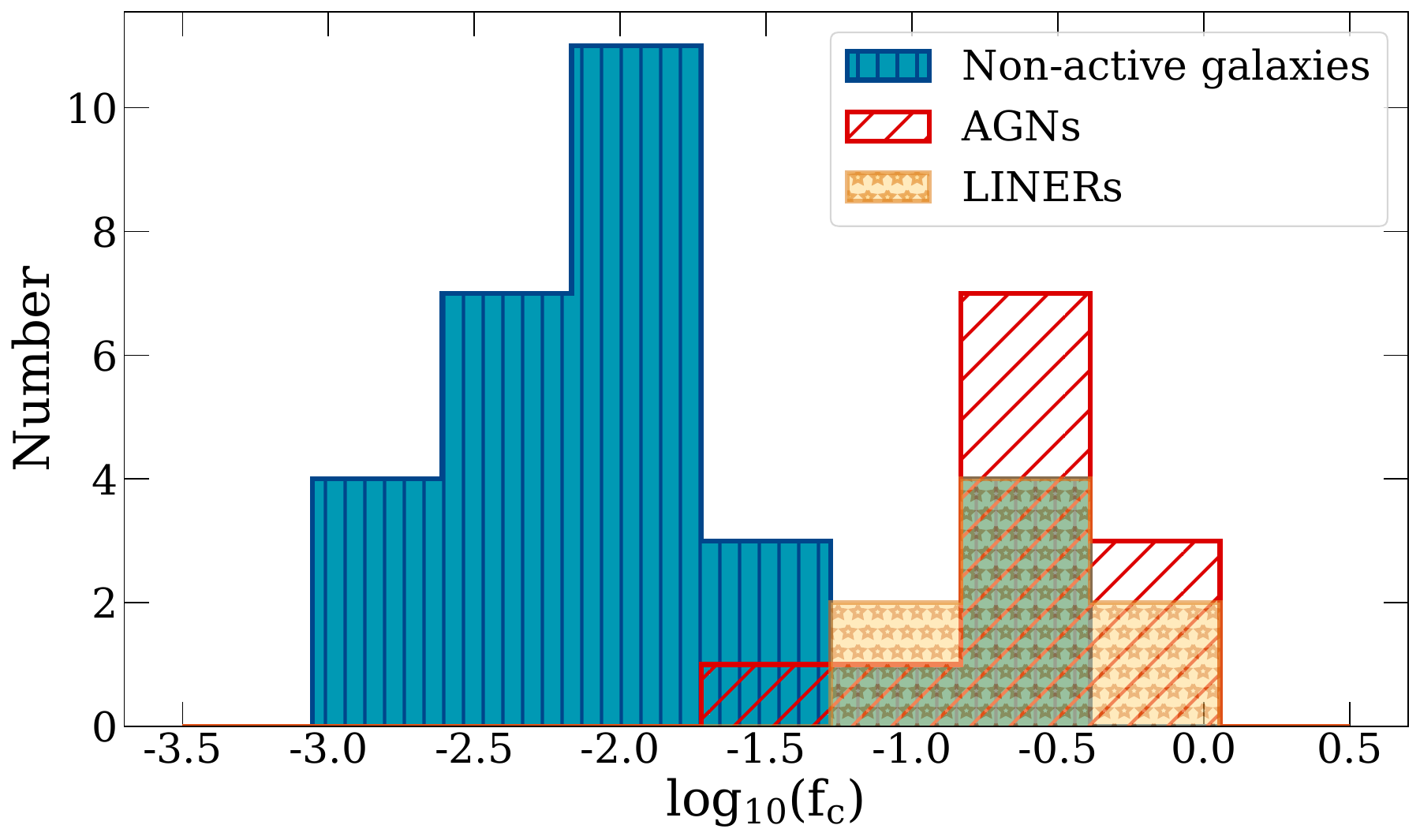}
  \caption{The covering factors distribution of non-active galaxies, AGNs, and LINERs.}\label{fig:covering_factor}
\end{figure}

In order to compare the dust covering factor ($f_c$) of AT2018dyk with other nuclear transients, we then collect those reported in literature before our work and acquire the values for 44 sources~\citep{yan2019,koll2020,jiang2021b,hinkle2022,wang2022}. They are calculated in a consistent way, that is the ratio of peak IR luminosity to the peak UV/optical luminosity. Moreover, we have calculated the $f_c$ of another five transients in LINERs reported by \cite{frederick2019} on our own, that is 0.633, 0.36, 0.298, 0.234, and 0.657 for ZTF18aasuray, ZTF18aahiqfi, ZTF18aaidlyq, ZTF18aaabltn, and ZTF18aasszwr, respectively.
Thus, the $f_c$ of a total of 50 sources have been obtained.
Among them, 20 sources occur in inactive galaxies, 12 in Seyferts, and 8 in LINERs according to the BPT classification of their host types. The distributions of dust covering factors for each type are shown in Figure \ref{fig:covering_factor}. 

The $f_c$ of non-active galaxies in our sample is $\sim0.01$ or even less with a median value of 0.01, that is at least one order of magnitude lower than that of AGNs (a median value of 0.23) consistent with \citet{jiang2021b}. It is notable that the $f_c$ of LINERs is comparable with that of AGNs, indicating a similar dusty structure for the two populations although their SMBH activities are hugely different. However, we can not rule out the possibility that the discussed LINERs undergoing a state transition belong to a special subclass, i.e., lying in the stage of intermittent accretion immediately after the Seyfert phase, unless they are actually TDEs such as in the case of AT2018dyk (see below).

\subsection{TDE scenario}
\subsubsection{The explanation of variability}
\cite{frederick2019} found that the optical spectra of AT2018dyk exhibited some characteristics, including blue continuum, and variations in some emission lines, such as H$\alpha$ and He \textsc{ii}. 
The light curves of AT2018dyk exhibit a sudden-rise and a power-law slow-decay profile in each band, which is similar to the case in TDEs. 
Moreover, the blackbody temperature fitted by UV/optical SED shows high ($\sim$20000~K) and no significant temporal evolution, which is consistent with the TDEs that have been found so far \citep{gezari2021}. Together with the broad emission lines in the optical spectra \citep{frederick2019} and soft X-ray, we suspect that it was induced by a TDE. In such a process, a star was disrupted by the SMBH in the system, which processes some prominent emission lines, resulting in the `changing-look' phenomenon. 

It remains possible that the activity of AGN can cause such a multi-wavelength outburst. Here, we can compare AT2018dyk with normal AGNs. The time scale of variability caused by disk instability is rather long (up to years) for such a supermassive black hole. Moreover, if the outburst is caused by disk instability, it will persist for quite some time after reaching peak luminosity \citep{zabludoff2021}. However, the variability time scale of AT2018dyk is on the order of months, and the UV/optical light curves decay after reaching peak luminosity, which does not resemble the activity of AGNs. Besides, the `bluer-when-brighter' trend can be seen in some AGNs, but AT2018dyk shows no significant color evolution. For X-ray spectra, normal AGNs can be fitted by a power-law model with a photon index ranging from 1.7 to 2.4, while in most X-ray detected TDEs, the photon index is commonly greater than or equal to 3 \citep{zabludoff2021}. In the case of AT2018dyk, the X-rays are rather soft, and the photon index is about 3 in the initial detection. These features suggest that AT2018dyk is not quite consistent with the activities of normal AGNs. Although most X-ray TDEs exhibit a thermal spectrum, exceptions have been found to date, such as XMMSL2 J144605.0+685735 \citep{saxton2019}. The X-ray spectrum of this source can be fitted well with a power-law model, which may be due to the black hole mass being so large ($\sim 7 \times 10^7 M_{\sun}$; \citealt{wevers2019}) that the temperature of the thermal radiation is too low to reach the X-ray band \citep{saxton2020}. Similarly to XMMSL2 J144605.0+685735, the black hole mass of AT2018dyk is also on the order of $10^7 M_{\sun}$, which may explain why the X-ray spectra can be fitted well by the soft power-law model.

It should be noted that \cite{frederick2019} reported the broadening of the Mg \textsc{ii} emission line as the UV/optical light curves decay, which they attributed to the delay of the light travel time of the Mg \textsc{ii} emission region. Although relatively common in quasars, the Mg \textsc{ii} emission line is rare in reported TDEs, although it has recently been detected in the newly discovered TDE candidate SDSS J014124+010306 \citep{zhang2022} and AT2021lwx \citep{wiseman2023,subrayan2023}. The absence of the Mg \textsc{ii} line may be related to the hot continuum \citep{cenko2016,brown2018,hung2019}. Interestingly, \cite{cenko2016} noted that the Mg \textsc{ii} absence phenomenon may be transient and can be seen as the continuum cools enough.

Most TDEs are currently found in optical surveys. The majority of these optically-discovered TDEs are not subsequently detected in X-ray follow-up observations \citep{gezari2021,hammerstein2023}. AT2018dyk is quite soft in the initial detections. As the flux increases, the X-ray hardens slightly. For reference, the X-ray in 1ES1927+654 exhibits a `harder-when-brighter' tendency in its rebrightening stage, and \cite{ricci2020} found that it may be caused by the reconstruction of the corona. Therefore, for AT2018dyk, the slightly hardening trend in the X-ray rising stage may also be related to the variation of the corona in the system, but its contribution may not be as obvious as in 1ES1927+654.  

During the optical peak, we estimate the UV/optical luminosity of $1.84\times10^{43}\rm erg/s$. Because no significant X-ray was detected at that time, we assume the bolometric luminosity equals the UV/optical luminosity. Its ratio to Eddington luminosity is $L_{\rm bol}/L_{Edd}=0.005$, where $L_{\rm Edd}$ is derived from the black hole mass $M_{\text{BH}}\sim 10^{7.38} M_{\sun}$ which we estimated by $M_{\rm BH}-\sigma$ relation. While for the X-ray peak, we assume $L_{\rm bol}$ as the sum of UV/optical and X-ray luminosity. Therefore, in the X-ray peak, we estimate the UV/optical luminosity is $L_{\rm UV/opt}=6.54\times10^{42}\rm erg/s$ and X-ray luminosity is $L_{\rm x}=4.21\times10^{42}\rm erg/s$, then we derive $L_{\rm bol}=L_{\rm UV/opt}+L_{\rm x}=1.75\times10^{43}\rm erg/s$, and $L_{\rm bol}/L_{\rm Edd}=0.005$. Compared with some reported TDEs, the luminosity of AT2018dyk is not as high. However, this may not be surprising because in TDEs, the peak accretion rate is influenced by the parameters of the SMBH and disrupted star,  with a relation of $\dot{M}_{\rm peak}/M_{\rm Edd}\propto M_6^{-3/2}m_{*}^2\beta^3r_{*}^{-3/2}$ where $M_6$ is the mass of the SMBH in the unit of $10^6M_{\sun}$, $m_*$ is the stellar mass, $r_*$ is the stellar radius and $\beta$ is the penetration factor \citep{lodato2011}. Given that the black hole mass is rather large, therefore, if a star is partially disrupted or has a large stellar radius could also result in a low accretion rate in TDE.

We integrate the UV/optical and X-ray luminosity and obtain that the energy released in these two ranges is $\sim4.1\times 10^{50}\text{erg}$ and $6.4\times 10^{50}~\text{erg}$, respectively. Summing the energy released in the X-ray, UV/optical, and IR bands, we derive a total energy of $\sim1.05\times10^{51}~\text{erg}$, which corresponds to an accreted mass of $0.01 M_{\sun}$ with an assumed accretion radiative efficiency of 0.1. This indicates that AT2018dyk is probably a partial TDE, like the cases of AT2019azh \citep{liu2022} and AT2019avd \citep{chen2022}.

\subsubsection{Time delay in the light curves}
The complete multiwavelength light curves exhibit a time delay for the X-ray and IR bands relative to the optical bands. Some TDEs have shown delayed X-ray emission, such as ASASSN-15oi \citep{gezari2017}, ASASSN-14li \citep{pasham2017}, OGLE16aaa \citep{shu2020}, AT2019azh \citep{liu2022}, and ATLAS17jrp \citep{wang2022}. Similar to these previously reported TDEs, AT2018dyk also has a delay in the X-ray that lags the $g$ band by $\sim 140$ days (peak-to-peak interval). The delayed X-ray emission may be related to the circularization of the debris stream, while the self-crossing of the stream produces the UV/optical emission, and the X-ray emission is only produced after disk formation \citep{piran2015}. However, assuming a solar-type star was disrupted by the SMBH, the circularization timescale would be $t_{\rm cir}=340.3~M_6^{-7/6}\beta^{-3}$ days \citep{gezari2017}. Considering the black hole mass of AT2018dyk, we find that its $t_{\rm cir}$ can range from a few days to tens of days depending on $\beta$. This is inconsistent with the observed X-ray delayed time. An alternative picture is that a newly accreted disk forms rapidly, but its inner region is obscured by some material that reprocesses the X-ray into the UV/optical bands. The X-ray photons cannot escape until the material becomes optically thin \citep{metzger2016} or moves away from the line of sight of the observer \citep{shu2020}. We estimate the timescale and find that the X-ray delayed time can be well explained by the escape time of the X-ray photons from the ionized material \citep{metzger2016} or the move time of the reprocessing layer \citep{shu2020}.

\subsubsection{The black hole mass and rising timescale}
After the star is tidally disrupted, the resulting debris stream falls back into the black hole in a timescale proportional to $M_{\rm BH}^{1/2}$ \citep{lodato2011,stone2013}. In addition to our initial ZTF-I sample \citep{lin2022}, we also collect some TDEs which are well sampled in the rising stage by other surveys, e.g., Pan-STARRs, PTF, etc. For the fitting of rising timescale, we select the $g$ band light curves unless $r$ band light curves are much better sampled. A Gaussian fit to the pre-peak luminosity and a redshift correction provides the rest-frame rising timescale $t_{\rm rise}$ for each source. Additionally, we collect central stellar velocity dispersion ($\sigma$) data from the literature for some host galaxies and estimate the black hole mass using the $M_{\rm BH}-\sigma$ relation of \cite{kormendy2013}. Notably, for AT2018dyk, we adopt $\sigma=112\pm4$ km/s \citep{french2020} and derive log${10}(M_{\rm BH}/M_{\sun})=7.38\pm0.31$, which is consistent with the result of \citet{frederick2019}, $M_{\rm BH}\sim 10^{7.6} M_{\sun}$.

\begin{figure}
  \centering
  \includegraphics[width=0.45\textwidth]{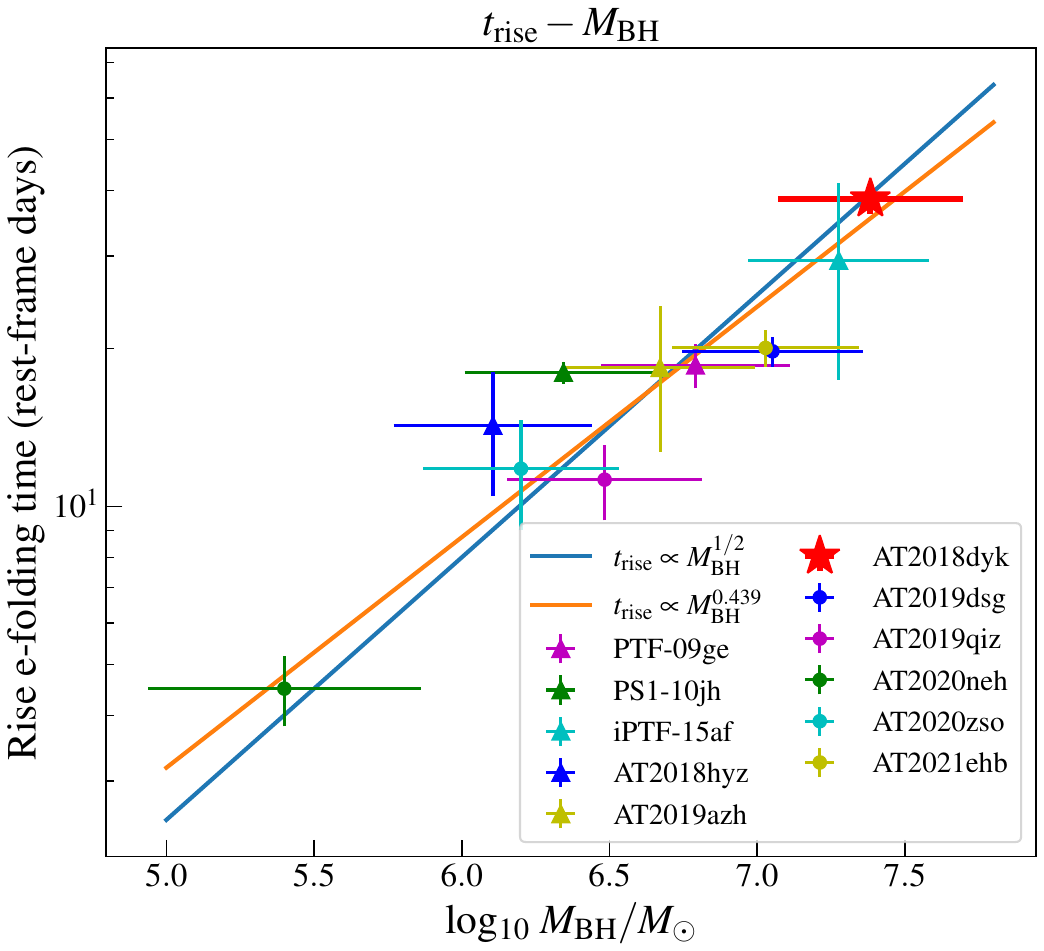}
  \caption{The correlation between the black hole mass ($M_{\rm BH}$) and the rest-frame rising timescale ($t_{\rm rise}$) of light curves. $M_{\rm BH}$ is estimated by the $M_{\rm BH}-\sigma$ relation of ~\protect\cite{kormendy2013}, where $\sigma$ represents the central stellar velocity dispersion for the hosts. We collect the $\sigma$ of PTF–09ge, PS1–10jh and iPTF–15af from ~\protect\cite{wevers2017}. Besides, we also collect the $\sigma$ of AT2018dyk \citep{french2020}, AT2018hyz \citep{short2020}, AT2019azh \citep{wevers2020}, AT2019dsg \citep{cannizzaro2021}, AT2019qiz \citep{nicholl2020}, AT2020neh \citep{angus2022}, AT2020zso \citep{wevers2022} and AT2021ehb \citep{yao2022}. $t_{\rm rise}$ for each source is derived from a Gaussian fit on the pre-peak luminosity and a redshift correction. The best fitting result for a power-law function of $t_{\rm rise}\propto M_{\rm BH}^\alpha$ is $\alpha=0.439\pm0.065$, which is plotted as the blue line. For reference, the theoretically predicted power-law function with an index of 0.5 is plotted in orange.}
  \label{fig:sigma_mass}
\end{figure}

Due to the limited $\sigma$ data and rising-stage light curves, only 11 sources (including AT2018dyk) are taken into final consideration. The correlation between $M_{\rm BH}$ and $t_{\rm rise}$ is shown in Figure \ref{fig:sigma_mass}. It is well consistent with the theoretical prediction, and also the correlation of $M_{\rm BH}$ and the fallback timescale reported by \cite{velzen2019,velzen2020} and \cite{yao2023}. We consider a power-law function of $t_{\rm rise}\propto M_{\rm BH}^\alpha$ for this correlation, and use \texttt{scipy.odr} to perform the fitting. The fitted index is $\alpha=0.439\pm0.065$. It is well consistent with the theoretical prediction ($\alpha=0.5$), and also the correlation of $M_{\rm BH}$ and the fallback timescale reported by \cite{velzen2019,velzen2020}. Given the facts that AT2018dyk is firmly in line with other reported TDEs, and the thermal, dynamic or viscous timescales for the accretion disk are all proportional to $M_{\rm BH}$, but not $M_{\rm BH}^{1/2}$ \citep{netzer2013}, we conclude that AT2018dyk should not be induced by the activity of the existed accretion disk, but probably a TDE. 

\section{Conclusion}\label{sec:conclusion}
AT2018dyk is a transient discovered in a LINER that exhibits bright emission in the X-ray, UV/optical, and IR bands. Its light curves display a `sudden-rise and slow-decay' profile that can be well fitted by a power-law function with an index of 1.58, close to the predicted trend by TDE theories. The X-ray spectrum is softer than typical AGNs during the outburst epoch, but a slight `harder-when-brighter' tendency can be seen before the flux reaches its peak, and a more prominent hardening trend is shown as the flux decreases. The outburst in IR bands indicates the presence of a dusty torus in the central region, which exhibits a decreasing dust temperature and a roughly constant radius as the IR luminosity declines. By fitting the IR data, we estimate a dust radius of $\sim10^{17}~\text{cm}$ and a dust covering factor of $\sim 0.4$. Our statistical analysis of the covering factors of the host galaxies where the 50 nuclear transients are located reveals that, in our sample, the covering factors of these LINERs are similar to AGNs rather than non-active galaxies, indicating the existence of dusty tori in these LINERs. We find a correlation between the rest-frame rising timescale for optical luminosity ($t_{\rm rise}$) and the black hole mass ($M_{\rm BH}$), $t_{\rm rise}\propto M_{\rm BH}^{1/2}$, for 10 reported TDEs, which is consistent with theoretical predictions. The fact that AT2018dyk also follows this correlation supports a TDE origin.

We propose that the delayed X-ray emission in AT2018dyk can be explained by a TDE. After a star is disrupted by an SMBH, the debris stream efficiently circularizes and forms a new accretion disk. However, the inner region of the disk is obscured by some material that reprocesses the X-ray into UV/optical bands, and the X-ray photons cannot escape until the material moves away or becomes optically thin.

Although more than 100 TDE candidates have been discovered, TDEs occurring around such massive black holes ($M_{\rm BH}\approx10^{7.38}M_{\odot}$) are still rare. As a transient in a LINER, an AT2018dyk-like source is an ideal probe to detect and measure the dusty torus. Moreover, our statistical results show that at least some of the LINERs host dusty tori, providing a plausible way to test the unified model of AGNs.

\section*{Acknowledgements}
We are grateful to the anonymous referee for providing valuable comments, which help to improve the quality of this work. This work is supported by the SKA Fast Radio Burst and High-Energy Transients Project (2022SKA0130102), the National Natural Science Foundation of China (grants 11833007, 12073025, 12192221) and the Fundamental Research Funds for the Central Universities (WK3440000006). The authors gratefully acknowledge the support of Cyrus Chun Ying Tang Foundations.
We acknowledge the use of public data from the Swift data archive. We thank the Swift ToO team for accepting our proposal and executing the observations. This work uses the observational data of XMM-Newton, an ESA science mission with instruments and contributions directly funded by ESA Member States and NASA. The ZTF forced-photometry service was funded under the Heising-Simons Foundation grant \#12540303 (PI: Graham).

\section*{Data Availability}

The data analyzed in this study can be freely retrieved from public archival database. The X-ray and UV/optical data are downloaded from the website of HEASARC website at \url{https://heasarc.gsfc.nasa.gov/db-perl/W3Browse/w3browse.pl}. ZTF and WISE photometry data are derived from the website \url{https://irsa.ipac.caltech.edu/}.

\bibliographystyle{mnras}

\bibliography{AT2018dyk}

\begin{thebibliography}{}
\makeatletter
\relax
\def\mn@urlcharsother{\let\do\@makeother \do\$\do\&\do\#\do\^\do\_\do\%\do\~}
\def\mn@doi{\begingroup\mn@urlcharsother \@ifnextchar [ {\mn@doi@}
  {\mn@doi@[]}}
\def\mn@doi@[#1]#2{\def\@tempa{#1}\ifx\@tempa\@empty \href
  {http://dx.doi.org/#2} {doi:#2}\else \href {http://dx.doi.org/#2} {#1}\fi
  \endgroup}
\def\mn@eprint#1#2{\mn@eprint@#1:#2::\@nil}
\def\mn@eprint@arXiv#1{\href {http://arxiv.org/abs/#1} {{\tt arXiv:#1}}}
\def\mn@eprint@dblp#1{\href {http://dblp.uni-trier.de/rec/bibtex/#1.xml}
  {dblp:#1}}
\def\mn@eprint@#1:#2:#3:#4\@nil{\def\@tempa {#1}\def\@tempb {#2}\def\@tempc
  {#3}\ifx \@tempc \@empty \let \@tempc \@tempb \let \@tempb \@tempa \fi \ifx
  \@tempb \@empty \def\@tempb {arXiv}\fi \@ifundefined
  {mn@eprint@\@tempb}{\@tempb:\@tempc}{\expandafter \expandafter \csname
  mn@eprint@\@tempb\endcsname \expandafter{\@tempc}}}

\bibitem[\protect\citeauthoryear{{Angus} et~al.,}{{Angus}
  et~al.}{2022}]{angus2022}
{Angus} C.~R.,  et~al., 2022, \mn@doi [Nature Astronomy]
  {10.1038/s41550-022-01811-y}, \href
  {https://ui.adsabs.harvard.edu/abs/2022NatAs...6.1452A} {6, 1452}

\bibitem[\protect\citeauthoryear{{Arcavi} et~al.,}{{Arcavi}
  et~al.}{2014}]{arcavi2014}
{Arcavi} I.,  et~al., 2014, \mn@doi [\apj] {10.1088/0004-637X/793/1/38}, \href
  {https://ui.adsabs.harvard.edu/abs/2014ApJ...793...38A} {793, 38}

\bibitem[\protect\citeauthoryear{{Arcavi} et~al.,}{{Arcavi}
  et~al.}{2018}]{arcavi2018}
{Arcavi} I.,  et~al., 2018, The Astronomer's Telegram, \href
  {https://ui.adsabs.harvard.edu/abs/2018ATel11953....1A} {11953, 1}

\bibitem[\protect\citeauthoryear{{Balmaverde} \& {Capetti}}{{Balmaverde} \&
  {Capetti}}{2015}]{balmaverde2015}
{Balmaverde} B.,  {Capetti} A.,  2015, \mn@doi [\aap]
  {10.1051/0004-6361/201526496}, \href
  {https://ui.adsabs.harvard.edu/abs/2015A&A...581A..76B} {581, A76}

\bibitem[\protect\citeauthoryear{{Blanchard} et~al.,}{{Blanchard}
  et~al.}{2017}]{blanchard2017}
{Blanchard} P.~K.,  et~al., 2017, \mn@doi [\apj] {10.3847/1538-4357/aa77f7},
  \href {https://ui.adsabs.harvard.edu/abs/2017ApJ...843..106B} {843, 106}

\bibitem[\protect\citeauthoryear{{Brown} et~al.,}{{Brown}
  et~al.}{2018}]{brown2018}
{Brown} J.~S.,  et~al., 2018, \mn@doi [\mnras] {10.1093/mnras/stx2372}, \href
  {https://ui.adsabs.harvard.edu/abs/2018MNRAS.473.1130B} {473, 1130}

\bibitem[\protect\citeauthoryear{{Burrows} et~al.,}{{Burrows}
  et~al.}{2005}]{burrows2005}
{Burrows} D.~N.,  et~al., 2005, \mn@doi [\ssr] {10.1007/s11214-005-5097-2},
  \href {https://ui.adsabs.harvard.edu/abs/2005SSRv..120..165B} {120, 165}

\bibitem[\protect\citeauthoryear{{Cannizzaro} et~al.,}{{Cannizzaro}
  et~al.}{2021}]{cannizzaro2021}
{Cannizzaro} G.,  et~al., 2021, \mn@doi [\mnras] {10.1093/mnras/stab851}, \href
  {https://ui.adsabs.harvard.edu/abs/2021MNRAS.504..792C} {504, 792}

\bibitem[\protect\citeauthoryear{{Cannizzaro}, {Levan}, {van Velzen}  \&
  {Brown}}{{Cannizzaro} et~al.}{2022}]{cannizzaro2022}
{Cannizzaro} G.,  {Levan} A.~J.,  {van Velzen} S.,   {Brown} G.,  2022, \mn@doi
  [\mnras] {10.1093/mnras/stac2014}, \href
  {https://ui.adsabs.harvard.edu/abs/2022MNRAS.516..529C} {516, 529}

\bibitem[\protect\citeauthoryear{{Cardelli}, {Clayton}  \& {Mathis}}{{Cardelli}
  et~al.}{1989}]{cardelli1989}
{Cardelli} J.~A.,  {Clayton} G.~C.,   {Mathis} J.~S.,  1989, \mn@doi [\apj]
  {10.1086/167900}, \href
  {https://ui.adsabs.harvard.edu/abs/1989ApJ...345..245C} {345, 245}

\bibitem[\protect\citeauthoryear{{Cenko} et~al.,}{{Cenko}
  et~al.}{2016}]{cenko2016}
{Cenko} S.~B.,  et~al., 2016, \mn@doi [\apjl] {10.3847/2041-8205/818/2/L32},
  \href {https://ui.adsabs.harvard.edu/abs/2016ApJ...818L..32C} {818, L32}

\bibitem[\protect\citeauthoryear{{Chen}, {Dou}  \& {Shen}}{{Chen}
  et~al.}{2022}]{chen2022}
{Chen} J.-H.,  {Dou} L.-M.,   {Shen} R.-F.,  2022, \mn@doi [\apj]
  {10.3847/1538-4357/ac558d}, \href
  {https://ui.adsabs.harvard.edu/abs/2022ApJ...928...63C} {928, 63}

\bibitem[\protect\citeauthoryear{{Falcke}, {K{\"o}rding}  \&
  {Markoff}}{{Falcke} et~al.}{2004}]{falcke2004}
{Falcke} H.,  {K{\"o}rding} E.,   {Markoff} S.,  2004, \mn@doi [\aap]
  {10.1051/0004-6361:20031683}, \href
  {https://ui.adsabs.harvard.edu/abs/2004A&A...414..895F} {414, 895}

\bibitem[\protect\citeauthoryear{{Frederick} et~al.,}{{Frederick}
  et~al.}{2019}]{frederick2019}
{Frederick} S.,  et~al., 2019, \mn@doi [\apj] {10.3847/1538-4357/ab3a38}, \href
  {https://ui.adsabs.harvard.edu/abs/2019ApJ...883...31F} {883, 31}

\bibitem[\protect\citeauthoryear{{French}, {Wevers}, {Law-Smith}, {Graur}  \&
  {Zabludoff}}{{French} et~al.}{2020}]{french2020}
{French} K.~D.,  {Wevers} T.,  {Law-Smith} J.,  {Graur} O.,   {Zabludoff}
  A.~I.,  2020, \mn@doi [\ssr] {10.1007/s11214-020-00657-y}, \href
  {https://ui.adsabs.harvard.edu/abs/2020SSRv..216...32F} {216, 32}

\bibitem[\protect\citeauthoryear{{Gezari}}{{Gezari}}{2021}]{gezari2021}
{Gezari} S.,  2021, \mn@doi [\araa] {10.1146/annurev-astro-111720-030029},
  \href {https://ui.adsabs.harvard.edu/abs/2021ARA&A..59...21G} {59, 21}

\bibitem[\protect\citeauthoryear{{Gezari}, {Cenko}  \& {Arcavi}}{{Gezari}
  et~al.}{2017}]{gezari2017}
{Gezari} S.,  {Cenko} S.~B.,   {Arcavi} I.,  2017, \mn@doi [\apjl]
  {10.3847/2041-8213/aaa0c2}, \href
  {https://ui.adsabs.harvard.edu/abs/2017ApJ...851L..47G} {851, L47}

\bibitem[\protect\citeauthoryear{{Gonz{\'a}lez-Mart{\'\i}n}
  et~al.,}{{Gonz{\'a}lez-Mart{\'\i}n} et~al.}{2015}]{martin2015}
{Gonz{\'a}lez-Mart{\'\i}n} O.,  et~al., 2015, \mn@doi [\aap]
  {10.1051/0004-6361/201425254}, \href
  {https://ui.adsabs.harvard.edu/abs/2015A&A...578A..74G} {578, A74}

\bibitem[\protect\citeauthoryear{{HI4PI Collaboration} et~al.,}{{HI4PI
  Collaboration} et~al.}{2016}]{HI4PI2016}
{HI4PI Collaboration} et~al., 2016, \mn@doi [\aap]
  {10.1051/0004-6361/201629178}, \href
  {https://ui.adsabs.harvard.edu/abs/2016A&A...594A.116H} {594, A116}

\bibitem[\protect\citeauthoryear{{Hammerstein} et~al.,}{{Hammerstein}
  et~al.}{2023}]{hammerstein2023}
{Hammerstein} E.,  et~al., 2023, \mn@doi [\apj] {10.3847/1538-4357/aca283},
  \href {https://ui.adsabs.harvard.edu/abs/2023ApJ...942....9H} {942, 9}

\bibitem[\protect\citeauthoryear{{Hickox} \& {Alexander}}{{Hickox} \&
  {Alexander}}{2018}]{hickox2018}
{Hickox} R.~C.,  {Alexander} D.~M.,  2018, \mn@doi [\araa]
  {10.1146/annurev-astro-081817-051803}, \href
  {https://ui.adsabs.harvard.edu/abs/2018ARA&A..56..625H} {56, 625}

\bibitem[\protect\citeauthoryear{{Hills}}{{Hills}}{1975}]{hills1975}
{Hills} J.~G.,  1975, \mn@doi [\nat] {10.1038/254295a0}, \href
  {https://ui.adsabs.harvard.edu/abs/1975Natur.254..295H} {254, 295}

\bibitem[\protect\citeauthoryear{{Hinkle}}{{Hinkle}}{2022}]{hinkle2022}
{Hinkle} J.~T.,  2022, arXiv e-prints, \href
  {https://ui.adsabs.harvard.edu/abs/2022arXiv221015681H} {p. arXiv:2210.15681}

\bibitem[\protect\citeauthoryear{{Hinkle}, {Holoien}, {Shappee}  \&
  {Auchettl}}{{Hinkle} et~al.}{2021}]{hinkle2021}
{Hinkle} J.~T.,  {Holoien} T. W.~S.,  {Shappee} B.~J.,   {Auchettl} K.,  2021,
  \mn@doi [\apj] {10.3847/1538-4357/abe4d8}, \href
  {https://ui.adsabs.harvard.edu/abs/2021ApJ...910...83H} {910, 83}

\bibitem[\protect\citeauthoryear{{Ho}}{{Ho}}{2008}]{ho2008}
{Ho} L.~C.,  2008, \mn@doi [\araa] {10.1146/annurev.astro.45.051806.110546},
  \href {https://ui.adsabs.harvard.edu/abs/2008ARA&A..46..475H} {46, 475}

\bibitem[\protect\citeauthoryear{{Holoien} et~al.,}{{Holoien}
  et~al.}{2022}]{holoien2022}
{Holoien} T. W.~S.,  et~al., 2022, \mn@doi [\apj] {10.3847/1538-4357/ac74b9},
  \href {https://ui.adsabs.harvard.edu/abs/2022ApJ...933..196H} {933, 196}

\bibitem[\protect\citeauthoryear{{Huang}, {Hu}, {Yin}, {Chen}, {Alexeeva},
  {Gao}  \& {Jiang}}{{Huang} et~al.}{2021}]{huang2021}
{Huang} S.,  {Hu} S.,  {Yin} H.,  {Chen} X.,  {Alexeeva} S.,  {Gao} D.,
  {Jiang} Y.,  2021, \mn@doi [\apj] {10.3847/1538-4357/ac0eff}, \href
  {https://ui.adsabs.harvard.edu/abs/2021ApJ...920...12H} {920, 12}

\bibitem[\protect\citeauthoryear{{Huang}, {Hu}, {Yin}, {Chen}, {Alexeeva}  \&
  {Jiang}}{{Huang} et~al.}{2022}]{huang2022}
{Huang} S.,  {Hu} S.,  {Yin} H.,  {Chen} X.,  {Alexeeva} S.,   {Jiang} Y.,
  2022, \mn@doi [\mnras] {10.1093/mnras/stac2022}, \href
  {https://ui.adsabs.harvard.edu/abs/2022MNRAS.515.2778H} {515, 2778}

\bibitem[\protect\citeauthoryear{{Hung} et~al.,}{{Hung}
  et~al.}{2019}]{hung2019}
{Hung} T.,  et~al., 2019, \mn@doi [\apj] {10.3847/1538-4357/ab24de}, \href
  {https://ui.adsabs.harvard.edu/abs/2019ApJ...879..119H} {879, 119}

\bibitem[\protect\citeauthoryear{{Jiang}, {Dou}, {Wang}, {Yang}, {Lyu}  \&
  {Zhou}}{{Jiang} et~al.}{2016}]{jiang2016}
{Jiang} N.,  {Dou} L.,  {Wang} T.,  {Yang} C.,  {Lyu} J.,   {Zhou} H.,  2016,
  \mn@doi [\apjl] {10.3847/2041-8205/828/1/L14}, \href
  {https://ui.adsabs.harvard.edu/abs/2016ApJ...828L..14J} {828, L14}

\bibitem[\protect\citeauthoryear{{Jiang} et~al.,}{{Jiang}
  et~al.}{2017}]{jiang2017}
{Jiang} N.,  et~al., 2017, \mn@doi [\apj] {10.3847/1538-4357/aa93f5}, \href
  {https://ui.adsabs.harvard.edu/abs/2017ApJ...850...63J} {850, 63}

\bibitem[\protect\citeauthoryear{{Jiang} et~al.,}{{Jiang}
  et~al.}{2021a}]{jiang2021}
{Jiang} N.,  et~al., 2021a, \mn@doi [\apjs] {10.3847/1538-4365/abd1dc}, \href
  {https://ui.adsabs.harvard.edu/abs/2021ApJS..252...32J} {252, 32}

\bibitem[\protect\citeauthoryear{{Jiang}, {Wang}, {Hu}, {Sun}, {Dou}  \&
  {Xiao}}{{Jiang} et~al.}{2021b}]{jiang2021b}
{Jiang} N.,  {Wang} T.,  {Hu} X.,  {Sun} L.,  {Dou} L.,   {Xiao} L.,  2021b,
  \mn@doi [\apj] {10.3847/1538-4357/abe772}, \href
  {https://ui.adsabs.harvard.edu/abs/2021ApJ...911...31J} {911, 31}

\bibitem[\protect\citeauthoryear{{Kool} et~al.,}{{Kool}
  et~al.}{2020}]{koll2020}
{Kool} E.~C.,  et~al., 2020, \mn@doi [\mnras] {10.1093/mnras/staa2351}, \href
  {https://ui.adsabs.harvard.edu/abs/2020MNRAS.498.2167K} {498, 2167}

\bibitem[\protect\citeauthoryear{{Kormendy} \& {Ho}}{{Kormendy} \&
  {Ho}}{2013}]{kormendy2013}
{Kormendy} J.,  {Ho} L.~C.,  2013, \mn@doi [\araa]
  {10.1146/annurev-astro-082708-101811}, \href
  {https://ui.adsabs.harvard.edu/abs/2013ARA&A..51..511K} {51, 511}

\bibitem[\protect\citeauthoryear{{Lin}, {Jiang}, {Kong}, {Huang}, {Lin}, {Zhu}
  \& {Wang}}{{Lin} et~al.}{2022}]{lin2022}
{Lin} Z.,  {Jiang} N.,  {Kong} X.,  {Huang} S.,  {Lin} Z.,  {Zhu} J.,   {Wang}
  Y.,  2022, \mn@doi [\apjl] {10.3847/2041-8213/ac9c63}, \href
  {https://ui.adsabs.harvard.edu/abs/2022ApJ...939L..33L} {939, L33}

\bibitem[\protect\citeauthoryear{{Liu}, {Dou}, {Chen}  \& {Shen}}{{Liu}
  et~al.}{2022}]{liu2022}
{Liu} X.-L.,  {Dou} L.-M.,  {Chen} J.-H.,   {Shen} R.-F.,  2022, \mn@doi [\apj]
  {10.3847/1538-4357/ac33a9}, \href
  {https://ui.adsabs.harvard.edu/abs/2022ApJ...925...67L} {925, 67}

\bibitem[\protect\citeauthoryear{{Lodato} \& {Rossi}}{{Lodato} \&
  {Rossi}}{2011}]{lodato2011}
{Lodato} G.,  {Rossi} E.~M.,  2011, \mn@doi [\mnras]
  {10.1111/j.1365-2966.2010.17448.x}, \href
  {https://ui.adsabs.harvard.edu/abs/2011MNRAS.410..359L} {410, 359}

\bibitem[\protect\citeauthoryear{{Mainzer} et~al.,}{{Mainzer}
  et~al.}{2014}]{Mainzer2014}
{Mainzer} A.,  et~al., 2014, \mn@doi [\apj] {10.1088/0004-637X/792/1/30}, \href
  {https://ui.adsabs.harvard.edu/abs/2014ApJ...792...30M} {792, 30}

\bibitem[\protect\citeauthoryear{{Masci} et~al.,}{{Masci}
  et~al.}{2019}]{masci2019}
{Masci} F.~J.,  et~al., 2019, \mn@doi [\pasp] {10.1088/1538-3873/aae8ac}, \href
  {https://ui.adsabs.harvard.edu/abs/2019PASP..131a8003M} {131, 018003}

\bibitem[\protect\citeauthoryear{{Mattila} et~al.,}{{Mattila}
  et~al.}{2018}]{mattila2018}
{Mattila} S.,  et~al., 2018, \mn@doi [Science] {10.1126/science.aao4669}, \href
  {https://ui.adsabs.harvard.edu/abs/2018Sci...361..482M} {361, 482}

\bibitem[\protect\citeauthoryear{{Metzger} \& {Stone}}{{Metzger} \&
  {Stone}}{2016}]{metzger2016}
{Metzger} B.~D.,  {Stone} N.~C.,  2016, \mn@doi [\mnras]
  {10.1093/mnras/stw1394}, \href
  {https://ui.adsabs.harvard.edu/abs/2016MNRAS.461..948M} {461, 948}

\bibitem[\protect\citeauthoryear{{Netzer}}{{Netzer}}{2013}]{netzer2013}
{Netzer} H.,  2013, {The Physics and Evolution of Active Galactic Nuclei}

\bibitem[\protect\citeauthoryear{{Nicholl} et~al.,}{{Nicholl}
  et~al.}{2020}]{nicholl2020}
{Nicholl} M.,  et~al., 2020, \mn@doi [\mnras] {10.1093/mnras/staa2824}, \href
  {https://ui.adsabs.harvard.edu/abs/2020MNRAS.499..482N} {499, 482}

\bibitem[\protect\citeauthoryear{{Pasham}, {Cenko}, {Sadowski}, {Guillochon},
  {Stone}, {van Velzen}  \& {Cannizzo}}{{Pasham} et~al.}{2017}]{pasham2017}
{Pasham} D.~R.,  {Cenko} S.~B.,  {Sadowski} A.,  {Guillochon} J.,  {Stone}
  N.~C.,  {van Velzen} S.,   {Cannizzo} J.~K.,  2017, \mn@doi [\apjl]
  {10.3847/2041-8213/aa6003}, \href
  {https://ui.adsabs.harvard.edu/abs/2017ApJ...837L..30P} {837, L30}

\bibitem[\protect\citeauthoryear{{Piran}, {Svirski}, {Krolik}, {Cheng}  \&
  {Shiokawa}}{{Piran} et~al.}{2015}]{piran2015}
{Piran} T.,  {Svirski} G.,  {Krolik} J.,  {Cheng} R.~M.,   {Shiokawa} H.,
  2015, \mn@doi [\apj] {10.1088/0004-637X/806/2/164}, \href
  {https://ui.adsabs.harvard.edu/abs/2015ApJ...806..164P} {806, 164}

\bibitem[\protect\citeauthoryear{{Ramos Almeida} \& {Ricci}}{{Ramos Almeida} \&
  {Ricci}}{2017}]{almeida2017}
{Ramos Almeida} C.,  {Ricci} C.,  2017, \mn@doi [Nature Astronomy]
  {10.1038/s41550-017-0232-z}, \href
  {https://ui.adsabs.harvard.edu/abs/2017NatAs...1..679R} {1, 679}

\bibitem[\protect\citeauthoryear{{Rees}}{{Rees}}{1988}]{rees1988}
{Rees} M.~J.,  1988, \mn@doi [\nat] {10.1038/333523a0}, \href
  {https://ui.adsabs.harvard.edu/abs/1988Natur.333..523R} {333, 523}

\bibitem[\protect\citeauthoryear{{Reynolds}, {Mattila}, {Efstathiou},
  {Kankare}, {Kool}, {Ryder}, {Pe{\~n}a-Mo{\~n}ino}  \&
  {P{\'e}rez-Torres}}{{Reynolds} et~al.}{2022}]{reynolds2022}
{Reynolds} T.~M.,  {Mattila} S.,  {Efstathiou} A.,  {Kankare} E.,  {Kool} E.,
  {Ryder} S.,  {Pe{\~n}a-Mo{\~n}ino} L.,   {P{\'e}rez-Torres} M.~A.,  2022,
  \mn@doi [\aap] {10.1051/0004-6361/202243289}, \href
  {https://ui.adsabs.harvard.edu/abs/2022A&A...664A.158R} {664, A158}

\bibitem[\protect\citeauthoryear{{Ricci} et~al.,}{{Ricci}
  et~al.}{2017a}]{ricci2017}
{Ricci} C.,  et~al., 2017a, \mn@doi [\apjs] {10.3847/1538-4365/aa96ad}, \href
  {https://ui.adsabs.harvard.edu/abs/2017ApJS..233...17R} {233, 17}

\bibitem[\protect\citeauthoryear{{Ricci} et~al.,}{{Ricci}
  et~al.}{2017b}]{ricci2017b}
{Ricci} C.,  et~al., 2017b, \mn@doi [\nat] {10.1038/nature23906}, \href
  {https://ui.adsabs.harvard.edu/abs/2017Natur.549..488R} {549, 488}

\bibitem[\protect\citeauthoryear{{Ricci} et~al.,}{{Ricci}
  et~al.}{2020}]{ricci2020}
{Ricci} C.,  et~al., 2020, \mn@doi [\apjl] {10.3847/2041-8213/ab91a1}, \href
  {https://ui.adsabs.harvard.edu/abs/2020ApJ...898L...1R} {898, L1}

\bibitem[\protect\citeauthoryear{{Roming} et~al.,}{{Roming}
  et~al.}{2005}]{roming2005}
{Roming} P. W.~A.,  et~al., 2005, \mn@doi [\ssr] {10.1007/s11214-005-5095-4},
  \href {https://ui.adsabs.harvard.edu/abs/2005SSRv..120...95R} {120, 95}

\bibitem[\protect\citeauthoryear{{Saxton} et~al.,}{{Saxton}
  et~al.}{2019}]{saxton2019}
{Saxton} R.~D.,  et~al., 2019, \mn@doi [\aap] {10.1051/0004-6361/201935650},
  \href {https://ui.adsabs.harvard.edu/abs/2019A&A...630A..98S} {630, A98}

\bibitem[\protect\citeauthoryear{{Saxton}, {Komossa}, {Auchettl}  \&
  {Jonker}}{{Saxton} et~al.}{2020}]{saxton2020}
{Saxton} R.,  {Komossa} S.,  {Auchettl} K.,   {Jonker} P.~G.,  2020, \mn@doi
  [\ssr] {10.1007/s11214-020-00708-4}, \href
  {https://ui.adsabs.harvard.edu/abs/2020SSRv..216...85S} {216, 85}

\bibitem[\protect\citeauthoryear{{Schlafly} \& {Finkbeiner}}{{Schlafly} \&
  {Finkbeiner}}{2011}]{schlafly2011}
{Schlafly} E.~F.,  {Finkbeiner} D.~P.,  2011, \mn@doi [\apj]
  {10.1088/0004-637X/737/2/103}, \href
  {https://ui.adsabs.harvard.edu/abs/2011ApJ...737..103S} {737, 103}

\bibitem[\protect\citeauthoryear{{Short} et~al.,}{{Short}
  et~al.}{2020}]{short2020}
{Short} P.,  et~al., 2020, \mn@doi [\mnras] {10.1093/mnras/staa2065}, \href
  {https://ui.adsabs.harvard.edu/abs/2020MNRAS.498.4119S} {498, 4119}

\bibitem[\protect\citeauthoryear{{Shu} et~al.,}{{Shu} et~al.}{2020}]{shu2020}
{Shu} X.,  et~al., 2020, \mn@doi [Nature Communications]
  {10.1038/s41467-020-19675-z}, \href
  {https://ui.adsabs.harvard.edu/abs/2020NatCo..11.5876S} {11, 5876}

\bibitem[\protect\citeauthoryear{{Stone}, {Sari}  \& {Loeb}}{{Stone}
  et~al.}{2013}]{stone2013}
{Stone} N.,  {Sari} R.,   {Loeb} A.,  2013, \mn@doi [\mnras]
  {10.1093/mnras/stt1270}, \href
  {https://ui.adsabs.harvard.edu/abs/2013MNRAS.435.1809S} {435, 1809}

\bibitem[\protect\citeauthoryear{{Subrayan} et~al.,}{{Subrayan}
  et~al.}{2023}]{subrayan2023}
{Subrayan} B.~M.,  et~al., 2023, \mn@doi [arXiv e-prints]
  {10.48550/arXiv.2302.10932}, \href
  {https://ui.adsabs.harvard.edu/abs/2023arXiv230210932S} {p. arXiv:2302.10932}

\bibitem[\protect\citeauthoryear{{Urry} \& {Padovani}}{{Urry} \&
  {Padovani}}{1995}]{urry1995}
{Urry} C.~M.,  {Padovani} P.,  1995, \mn@doi [\pasp] {10.1086/133630}, \href
  {https://ui.adsabs.harvard.edu/abs/1995PASP..107..803U} {107, 803}

\bibitem[\protect\citeauthoryear{{Wang} et~al.,}{{Wang}
  et~al.}{2022}]{wang2022}
{Wang} Y.,  et~al., 2022, \mn@doi [\apjl] {10.3847/2041-8213/ac6670}, \href
  {https://ui.adsabs.harvard.edu/abs/2022ApJ...930L...4W} {930, L4}

\bibitem[\protect\citeauthoryear{{Wevers}}{{Wevers}}{2020}]{wevers2020}
{Wevers} T.,  2020, \mn@doi [\mnras] {10.1093/mnrasl/slaa097}, \href
  {https://ui.adsabs.harvard.edu/abs/2020MNRAS.497L...1W} {497, L1}

\bibitem[\protect\citeauthoryear{{Wevers}, {van Velzen}, {Jonker}, {Stone},
  {Hung}, {Onori}, {Gezari}  \& {Blagorodnova}}{{Wevers}
  et~al.}{2017}]{wevers2017}
{Wevers} T.,  {van Velzen} S.,  {Jonker} P.~G.,  {Stone} N.~C.,  {Hung} T.,
  {Onori} F.,  {Gezari} S.,   {Blagorodnova} N.,  2017, \mn@doi [\mnras]
  {10.1093/mnras/stx1703}, \href
  {https://ui.adsabs.harvard.edu/abs/2017MNRAS.471.1694W} {471, 1694}

\bibitem[\protect\citeauthoryear{{Wevers} et~al.,}{{Wevers}
  et~al.}{2019}]{wevers2019}
{Wevers} T.,  et~al., 2019, \mn@doi [\mnras] {10.1093/mnras/stz1602}, \href
  {https://ui.adsabs.harvard.edu/abs/2019MNRAS.487.4136W} {487, 4136}

\bibitem[\protect\citeauthoryear{{Wevers} et~al.,}{{Wevers}
  et~al.}{2022}]{wevers2022}
{Wevers} T.,  et~al., 2022, \mn@doi [\aap] {10.1051/0004-6361/202142616}, \href
  {https://ui.adsabs.harvard.edu/abs/2022A&A...666A...6W} {666, A6}

\bibitem[\protect\citeauthoryear{{Wiseman} et~al.,}{{Wiseman}
  et~al.}{2023}]{wiseman2023}
{Wiseman} P.,  et~al., 2023, \mn@doi [\mnras] {10.1093/mnras/stad1000}, \href
  {https://ui.adsabs.harvard.edu/abs/2023MNRAS.522.3992W} {522, 3992}

\bibitem[\protect\citeauthoryear{{Wright} et~al.,}{{Wright}
  et~al.}{2010}]{Wright2010}
{Wright} E.~L.,  et~al., 2010, \mn@doi [\aj] {10.1088/0004-6256/140/6/1868},
  \href {https://ui.adsabs.harvard.edu/abs/2010AJ....140.1868W} {140, 1868}

\bibitem[\protect\citeauthoryear{{Yan} et~al.,}{{Yan} et~al.}{2019}]{yan2019}
{Yan} L.,  et~al., 2019, \mn@doi [\apj] {10.3847/1538-4357/ab074b}, \href
  {https://ui.adsabs.harvard.edu/abs/2019ApJ...874...44Y} {874, 44}

\bibitem[\protect\citeauthoryear{{Yao} et~al.,}{{Yao} et~al.}{2022}]{yao2022}
{Yao} Y.,  et~al., 2022, \mn@doi [\apj] {10.3847/1538-4357/ac898a}, \href
  {https://ui.adsabs.harvard.edu/abs/2022ApJ...937....8Y} {937, 8}

\bibitem[\protect\citeauthoryear{{Yao} et~al.,}{{Yao} et~al.}{2023}]{yao2023}
{Yao} Y.,  et~al., 2023, \mn@doi [arXiv e-prints] {10.48550/arXiv.2303.06523},
  \href {https://ui.adsabs.harvard.edu/abs/2023arXiv230306523Y} {p.
  arXiv:2303.06523}

\bibitem[\protect\citeauthoryear{{Zabludoff} et~al.,}{{Zabludoff}
  et~al.}{2021}]{zabludoff2021}
{Zabludoff} A.,  et~al., 2021, \mn@doi [\ssr] {10.1007/s11214-021-00829-4},
  \href {https://ui.adsabs.harvard.edu/abs/2021SSRv..217...54Z} {217, 54}

\bibitem[\protect\citeauthoryear{{Zhang}}{{Zhang}}{2022}]{zhang2022}
{Zhang} X.-G.,  2022, \mn@doi [\mnras] {10.1093/mnrasl/slac092}, \href
  {https://ui.adsabs.harvard.edu/abs/2022MNRAS.516L..66Z} {516, L66}

\bibitem[\protect\citeauthoryear{{Zhang} et~al.,}{{Zhang}
  et~al.}{2022}]{zhangw2022}
{Zhang} W.~J.,  et~al., 2022, \mn@doi [\aap] {10.1051/0004-6361/202142253},
  \href {https://ui.adsabs.harvard.edu/abs/2022A&A...660A.119Z} {660, A119}

\bibitem[\protect\citeauthoryear{{van Velzen}, {Mendez}, {Krolik}  \&
  {Gorjian}}{{van Velzen} et~al.}{2016}]{velzen2016}
{van Velzen} S.,  {Mendez} A.~J.,  {Krolik} J.~H.,   {Gorjian} V.,  2016,
  \mn@doi [\apj] {10.3847/0004-637X/829/1/19}, \href
  {https://ui.adsabs.harvard.edu/abs/2016ApJ...829...19V} {829, 19}

\bibitem[\protect\citeauthoryear{{van Velzen}, {Stone}, {Metzger}, {Gezari},
  {Brown}  \& {Fruchter}}{{van Velzen} et~al.}{2019}]{velzen2019}
{van Velzen} S.,  {Stone} N.~C.,  {Metzger} B.~D.,  {Gezari} S.,  {Brown}
  T.~M.,   {Fruchter} A.~S.,  2019, \mn@doi [\apj] {10.3847/1538-4357/ab1844},
  \href {https://ui.adsabs.harvard.edu/abs/2019ApJ...878...82V} {878, 82}

\bibitem[\protect\citeauthoryear{{van Velzen}, {Holoien}, {Onori}, {Hung}  \&
  {Arcavi}}{{van Velzen} et~al.}{2020}]{velzen2020}
{van Velzen} S.,  {Holoien} T. W.~S.,  {Onori} F.,  {Hung} T.,   {Arcavi} I.,
  2020, \mn@doi [\ssr] {10.1007/s11214-020-00753-z}, \href
  {https://ui.adsabs.harvard.edu/abs/2020SSRv..216..124V} {216, 124}

\makeatother
\end{thebibliography}
\end{document}